\documentclass[12pt,a4paper] {article}
\usepackage{epsfig,graphics,amssymb}

\newcommand{\be}{\begin{equation}}
\newcommand{\ee}{\end{equation}}
\newcommand{\bear}{\begin{eqnarray}}
\newcommand{\ear}{\end{eqnarray}}
\newcommand{\bea}{\begin{eqnarray*}}
\newcommand{\ea}{\end{eqnarray*}}

\newcommand{\simgt}{\hbox{ \raise3pt\hbox to 0pt{$>$}
    \raise-3pt\hbox{$\sim$} }}
\newcommand{\simsm}{\hbox{ \raise3pt\hbox to 0pt{$<$}
    \raise-3pt\hbox{$\sim$} }}

\begin{document}
\begin{flushright}
DAMTP-1998-99\\
HD-THEP-98-28
\end{flushright}
\vspace{3cm}

\begin{center}
{\LARGE
Some Remarks on the Pomeron and the Odderon}\\
\bigskip
{\LARGE in Theory
and Experiment}\\
\vspace{2cm}

{\sc P. V. Landshoff
\footnote{P.V.Landshoff@damtp.cambridge.ac.uk}}\\
\medskip
{\em DAMTP, University of Cambridge, U.K.}\\

\vspace{.5cm}
and \\
\vspace{.5cm}
{\sc O. Nachtmann
\footnote{O.Nachtmann@thphys.uni-heidelberg.de}
}
\\
\medskip
{\em Institut f{\"u}r Theoretische Physik} \\
{\em Universit\"at Heidelberg, FRG}

\end{center}
\vspace{3cm}

\begin{abstract}
On March 19-21, 1998, a workshop devoted to questions of the pomeron
and the odderon in high energy scattering was held in Heidelberg.
This note gives a personal
account of some of the issues discussed at
this workshop. Of course,
misconceptions and misunderstandings
are to be blamed on us, not on the other participants
of the workshop. A puzzle of odderon physics is identified and a convenient
reaction for its experimental study is discussed.
\end{abstract}

\newpage
\noindent{\bf The pomeron}

The pomeron as an effective object whose exchange governs
high energy diffractive reactions is well established phenomenologically.
It carries vacuum quantum numbers, $C=P=+1$. A very successful
theoretical description of diffractive hadron-hadron reactions
in terms of Regge language \cite{1}
was developed in \cite{1a}. How to derive these pomeron effects
in the framework of QCD is still a problem lacking a complete
solution \cite{1b}.
The basic suggestion was made more than 20 years ago in \cite{2}:
In the simplest picture the pomeron is some sort of two-gluon
exchange.

With the HERA discovery of the rapid rise of the structure function
at small $x$,  and the
observation of ``hard'' diffractive reactions first at the CERN
collider\cite{ua8} and then
at HERA \cite{3},  the situation has changed dramatically.
We now talk of a ``soft'' pomeron,  responsible for example for
the slow rise with energy of hadron-hadron total cross-sections,
and a ``hard'' pomeron, revealed first in the behaviour
of the proton structure function at small $x$.
This immediately raises the question whether there are two separate
pomerons \cite{4}, or whether instead the soft pomeron becomes progressively
harder as $Q^2$ increases \cite{capella}.

The controversy in the literature about this question is related to another
matter of disagreement: whether the effect of single-pomeron exchange is  
significantly reduced by multiple exchange. Certainly, such ``shadowing''
must occur at some level, but the magnitude of its effect on
total cross-sections cannot be calculated. The simplest assumption is that it
is rather small, and then one explains the apparent hardening
of the pomeron with increasing $Q^2$ by saying that the soft and
hard pomerons are separate, with the relative size of their contributions
changing with $Q^2$. On the other hand, it could also be that there
is a single pomeron which is rather hard, but that in soft processes
the shadowing is sufficient to reduce its apparent hardness \cite{8a}.

As discussed above, the transition from the soft to the hard
pomeron regime has been studied extensively in $ep$ scattering.
Another interesting reaction for such studies is vector meson
production in two photon processes at $e^+e^-$ colliders
(Fig. 1)
\bear\label{1a}
&&\gamma(q_1)+\gamma(q_2)\longrightarrow V_1(p_1)+V_2(p_2)\nonumber\\
&&V_1,V_2=\rho^0,\omega,\phi,J/\psi... .\ear
At least to some extent the initial photons will act like hadrons in
the sense of the vector dominance model. In (\ref{1a}) we
can have small or large
virtualities $|q^2_1|,|q^2_2|$ of the photons
in the initial state and light or heavy vector mesons
in the final state. It will be very revealing to see which
combinations of the initial and final states
correspond to either the soft or the hard pomeron regime.
For theoretical work on this question we refer to \cite{8b}.
An experimental study of the reaction (\ref{1a}) at least
for $V_1=V_2=\rho^0$ should be feasible at LEP2 \cite{8c}.

Until very recently, there was good reason to believe that,
while the soft pomeron is surely nonperturbative and therefore
very difficult to analyse with the theoretical techniques at
present available,  the properties of the hard pomeron can be calculated
from perturbative QCD.
In suitable hard processes we can undoubtedly have reactions
where the exchange of two perturbative gluons
having together $C=P=+1$ dominates. Simple two-gluon exchange
gives a Regge trajectory $\alpha_{{\rm l}\!{\rm P}}(t)=1$.
Higher order corrections to such
an exchange in perturbative QCD (pQCD) should under suitable
conditions lead to effects associated with the so-called
``Lipatov pomeron'' \cite{5, 1b}. Up to this year these corrections
were estimated to be very large, changing the
intercept from
1 to $\approx 1.4$. More recent calculations, on the other hand,
predict a very small, perhaps even negative, change of intercept \cite{6}.
Ways to look for effects typical of these pQCD radiative
corrections connected with the Lipatov pomeron have been
discussed by many authors \cite{7}.

An approach to the theory underlying the soft pomeron has been developed
in \cite{8,9,10,11}. The nonperturbative
QCD-based model which gives a microscopic
description of the soft pomeron is indeed quite successful in
comparison with experiment on hadron-hadron elastic scattering
\cite{10,11,12}
and a number of other reactions \cite{13}. A very similar approach
to hard diffractive phenomena as observed at HERA was developed in
\cite{14}. How to relate this approach to the Regge language
is still unclear, but there are interesting ideas in this
direction \cite{15}.

\bigskip
\noindent{\bf The odderon}

Now we come to the odderon, which was introduced as the
$C=P=-1$ partner of the pomeron in terms of Regge language
in \cite{16}. Let us start with the perturbative regime. Undoubtedly
there are reactions which are dominated in a suitable kinematic regime
by the exchange of 3 perturbative gluons carrying together $C=P=-1$. An
example where this is believed to be the case is large
angle $pp$ scattering \cite{17} with the dominant diagrams indicated
in Fig. 2.

Clearly this perturbative odderon, the partner of the
perturbative two-gluon pomeron, exists and it has been observed
experimentally. As with the pomeron, the question  arises of pQCD corrections
to this type of odderon. A lot of theoretical work has
been devoted to this \cite{18}-\cite{21} and
the result seems to be that such corrections have a small
effect changing for instance the (effective) intercept of the
odderon trajectory $\alpha_{\mathbb{O}}(0)$ from 1 for the
``naked'' 3 gluon exchange by less than 10 \%
\cite{19,21,22}. Thus
our conclusion is that -- as for the pomeron case -- it will be hard
to establish experimentally for this type of odderon the particular pQCD effects
connected with the higher-orders summation in the spirit of Lipatov.

It is our opinion that effects of $C=P=-1$ exchange in soft
high-energy hadronic reactions will again involve the
nonperturbative features of QCD in an essential way
\cite{9,11}. This type of odderon is the one introduced originally
in terms of Regge language and gives rise for instance
to a difference between the amplitudes of $pp$ and
$p\bar p$ elastic scattering  at $t=0$. Such a difference has
indeed been seen \cite{dip} in the dip region in high-energy $pp$
and $p\bar p$ elastic scattering,  but at $t=0$
data on $p\bar p$ for $\sqrt s\stackrel{\scriptstyle>}{\sim}0.5$ TeV
\cite{23},  together with dispersion relations instead of the as-%
yet unavailable $pp$ data at similar energies, are usually interpreted
as giving tight bounds for such a difference:
\be\label{2}
|\rho_{pp}(s)-\rho_{p\bar p}(s)|\stackrel{\scriptstyle<}{\sim}
0.05,\ee
where
\be\label{3}
\rho(s)=\frac{{\rm Re}\ {\cal T}(s,t)}{{\rm Im}\ {\cal T}(s,t)}\Bigg|_{t=0}.
\ee
For a different view see \cite{24}.

For us, the puzzle of odderon physics is: {\sl
why has the soft odderon not been observed so far
at $t=0$?} Various
suggestions to explain this fact have been made
\cite{19,20,25,26}. In \cite{27} it was argued that in
the framework of the model for high energy diffractive reactions of
\cite{9,10} the odderon should not couple in elastic
meson-meson, meson-baryon and meson-antibaryon scattering
\bear\label{4}
&&M\ +\ M\ \longrightarrow\ M\ +\ M\nonumber\\
&&M\ +\ B\ \longrightarrow\ M\ +\ B\nonumber\\
&&M\ +\ \bar B\ \longrightarrow\ M\ +\ \bar B\ear
and also not in baryon-baryon and baryon-antibaryon
scattering
\bear\label{5}
&&B\ +\ B\ \longrightarrow\ B\ +\ B\nonumber\\
&&B\ +\ \bar B\ \longrightarrow\ B\ +\ \bar B\ear
if the baryons have a spatial linear structure,
consisting of a quark and a diquark. For baryons where the
three valence quarks are well separated in star-like
configurations, {large} odderon effects
are predicted. Thus, in this model the {soft odderon}
effects are related to the {internal baryon structure}.
Large effects from the soft odderon are predicted for inelastic
diffractive processes, for example double diffractive break-up,
\be\label{6}
B_1\ +\ B_2\ \longrightarrow\ B_1^*\ +\ B^*_2\ee
where $B^*_1$ and $B^*_2$ stand for diffractively excited baryons and
for continuum states (Fig. 3).

A particularly convenient  reaction where all aspects of odderon
physics could be studied seems to be exclusive $C=+1$ meson
production in $ep$ collisions \cite{28,29}, both without and
with diffractive proton breakup:
\bear\label{7}
&&e\ +\ p\ \longrightarrow\ e\ +\ M\ +\ p,\nonumber\\
&&e\ +\ p\ \longrightarrow\ e\ +\ M\ +\ N^*.\ear

Here both $\gamma \mathbb{O}$ and $\gamma\gamma$
exchange can contribute
(Fig. 4).
For $Q^2=-q^2\stackrel{\scriptstyle<}{\sim}0.5\ GeV^2$
and light mesons $M=\pi^0,\ \eta,\ \eta',\ f(1270)$,
we expect to have  ``soft'' odderon exchange. The prediction of
\cite{27,30} is that, if the proton has a quark-diquark structure,
the amplitude should be small for
the elastic case, i.e. for $p$ in the final state.
On the other hand, a large
amplitude is predicted for diffractive breakup, for instance
for $N(1535)$ (a state with $J^P=1/2^-$) production. If we go now to
heavier mesons, $M=\eta_c,\eta_b$, and/or increase $Q^2$
beyond $1\ GeV^2$, we should come into the perturbative
regime where the reactions (\ref{7}) should be dominated by
the exchange of three perturbative gluons \cite{31}. An
additional bonus for the reactions (\ref{7}) is that the interference
of the $\gamma \mathbb{O}$ with the $\gamma\gamma$ exchange
amplitude allows one to get information on the phase of the
odderon exchange amplitude. Thus, our conclusion is that it should
be very worthwhile to study the reactions (\ref{7}) experimentally
and this can indeed be done at HERA \cite{32}.

\bigskip
\noindent
{\bf Acknowledgements}

Our thanks are due to all participants of the workshop held
in Heidelberg in March 98 for their clear talks and lively
discussions which helped us to come to the views expressed
in our paper. In particular we want to thank the speakers
E. Berger, M. Erdmann, A. Donnachie, T. Gousset, D. Graudenz,
A. Hebecker, W. Kilian, R. Kirschner, J. Kwiecinski, S. V. Levonian,
L. Lindemann, G. Matthiae, E. Meggiolaro, B. Nicolescu,
M. R\"uter, P. Schlein, S. S\"oldner-Rembold, S. Tapprogge, and
D. Westphal. We also want to thank A. Donnachie and H. G. Dosch
for a critical reading of the manuscript and C. Schwanenberger for help
with the figures.

This research was supported by the British Council and the German
Academic Exchange Service (ARC projects 577 and 313), by the
EC Programme ``Training and Mobility of Researchers'', Network
``Hadronic Physics with High Energy Electromagnetic Probes'',
contract ERB FMRX-CT96-0008, and by PPARC.

\newpage

\noindent{\bf Figure Captions}\\

\bigskip
\noindent Fig. 1: Vector meson production in two
photon processes at $e^+e^-$ colliders

\bigskip
\noindent Fig. 2: Three gluon exchange in large angle $pp$ scattering

\bigskip
\noindent
Fig. 3: Double diffractive breakup in baryon-baryon collisions with
odderon exchange

\bigskip
\noindent
Fig. 4: $C=+1$ meson production in $ep$ collisions with elastic or
inelastic proton scattering

\newpage
\pagestyle{empty}

\begin{figure}[h]
  \begin{center}
    \epsfig{file=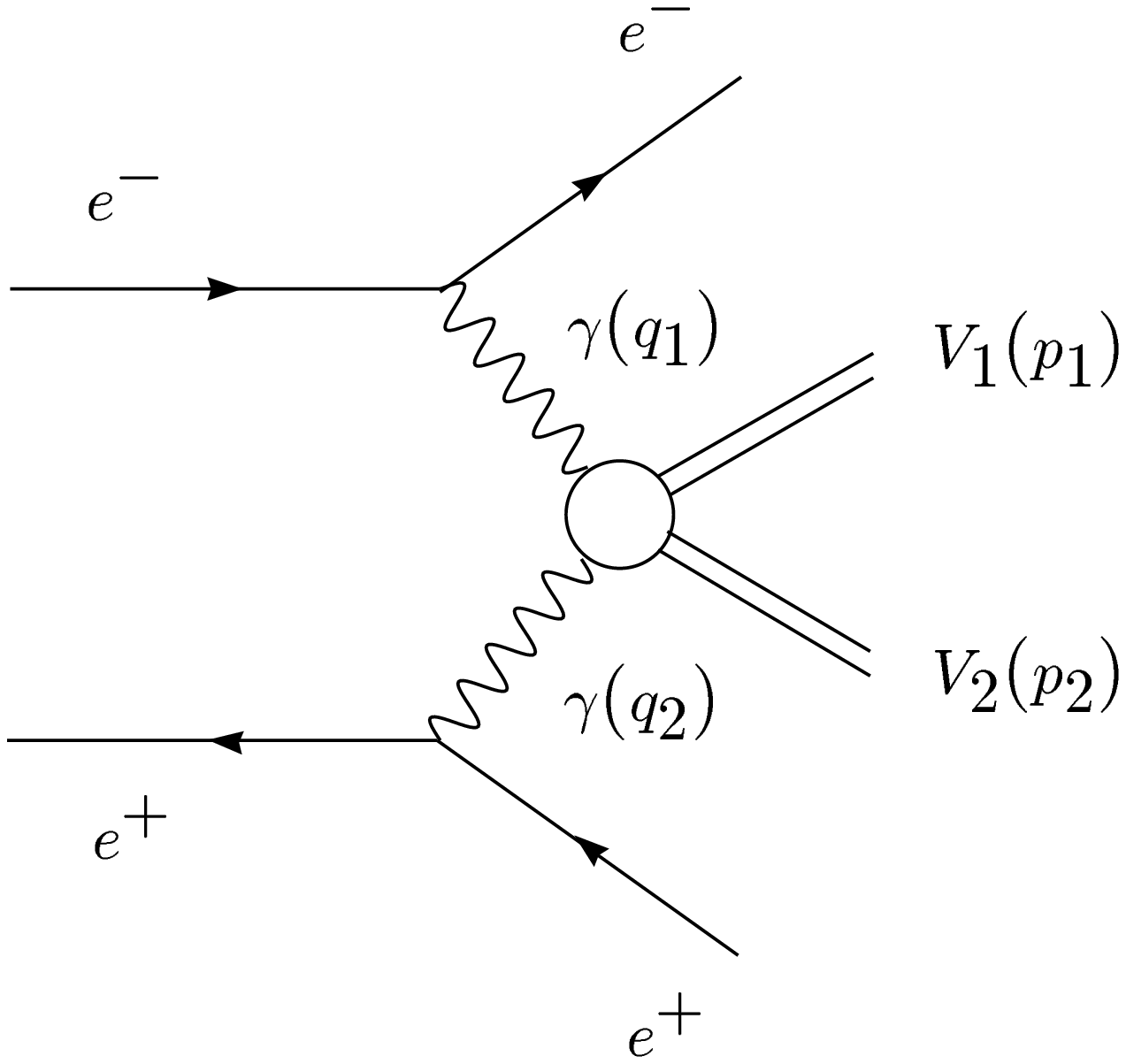,width=0.6\hsize}
  \end{center}
\label{fig:pic1}
\end{figure}

\begin{center}
  {\bf Fig. 1}
\end{center}
\newpage

\begin{figure}[t]
  \begin{center}
    \epsfig{file=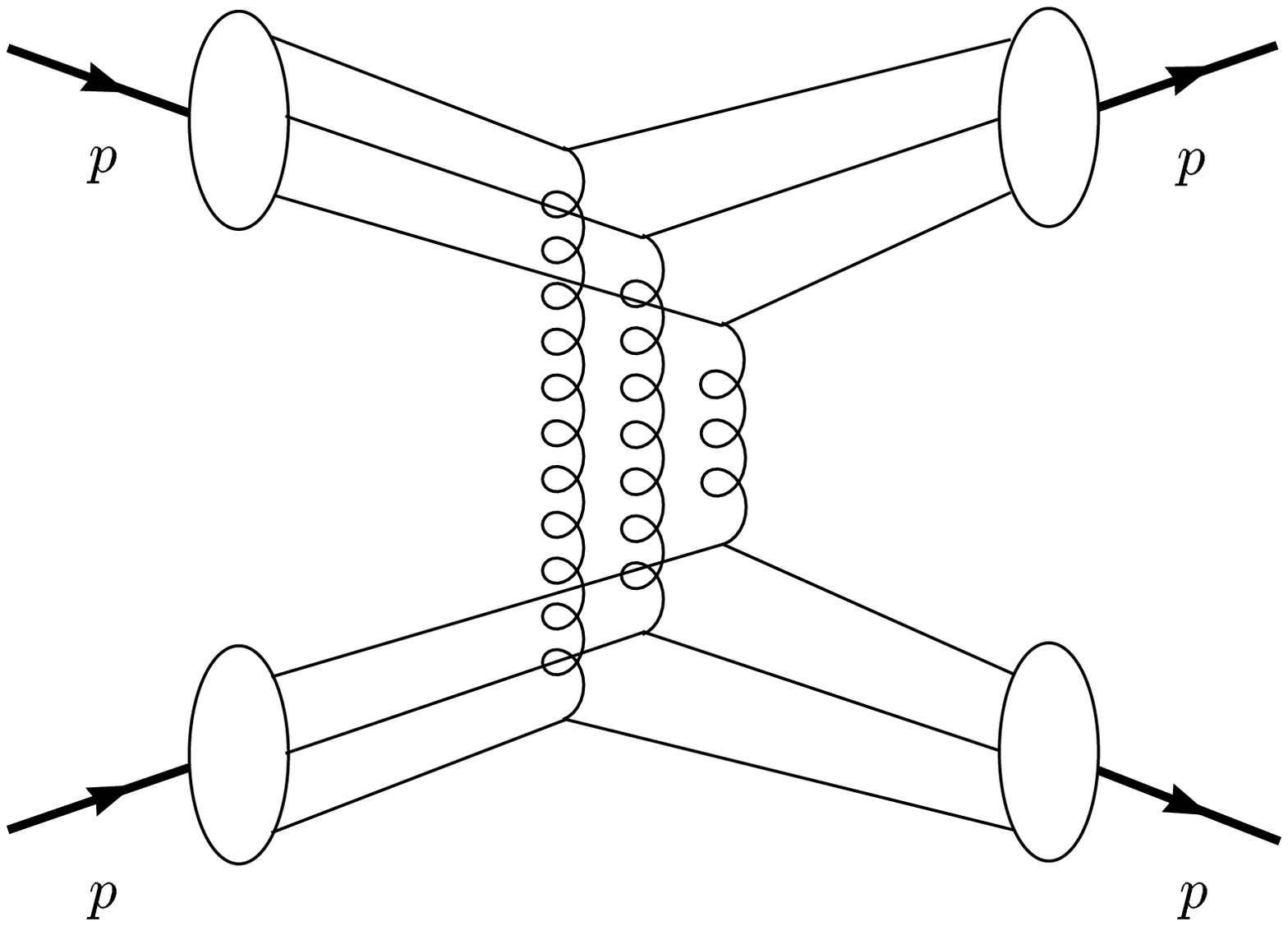,width=0.6\hsize}
  \end{center}
\label{fig:pic2}
\end{figure}

\begin{center}
  {\bf Fig. 2}
\end{center}
\vspace{3cm}

\begin{figure}[h]
  \begin{center}
    \epsfig{file=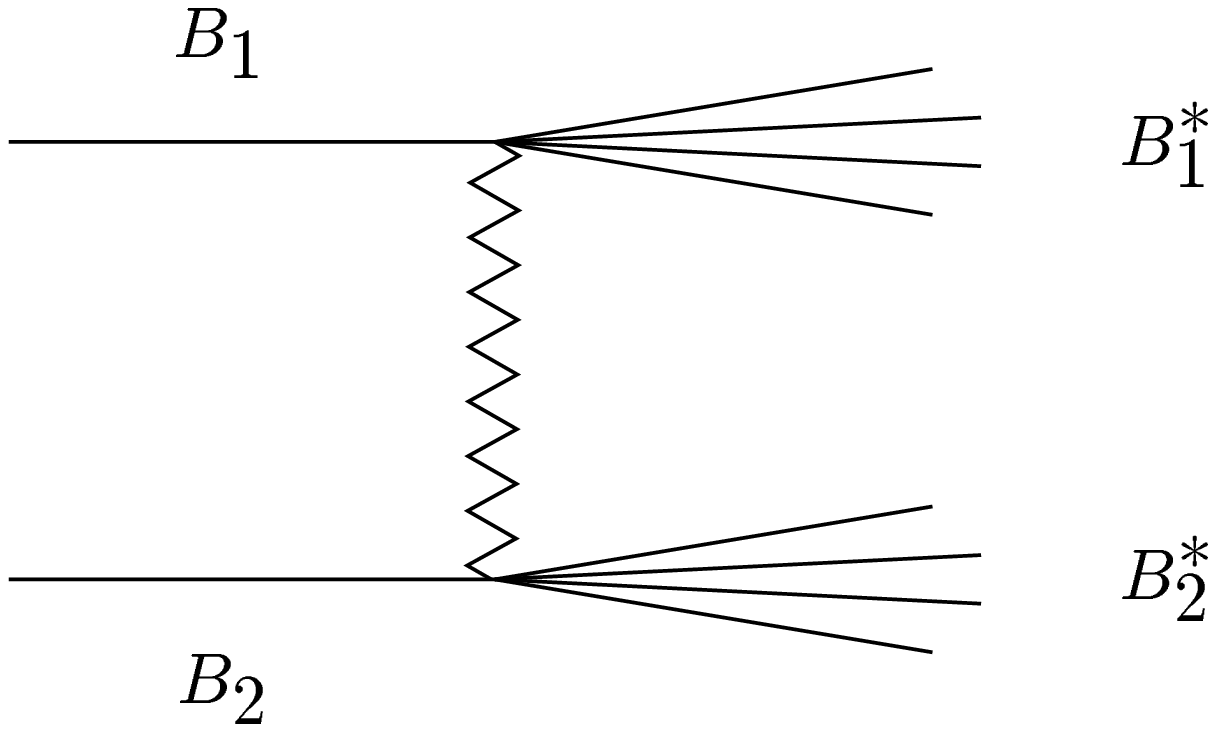,width=0.6\hsize}
  \end{center}
\vspace{-3.7cm}
\hspace{7.7cm}
{\LARGE $\mathbb{O}$}
\vspace{3.9cm}
\label{fig:pic3}
\end{figure}

\begin{center}
  {\bf Fig. 3}
\end{center}
\newpage

\begin{figure}[t]
  \begin{center}
    \epsfig{file=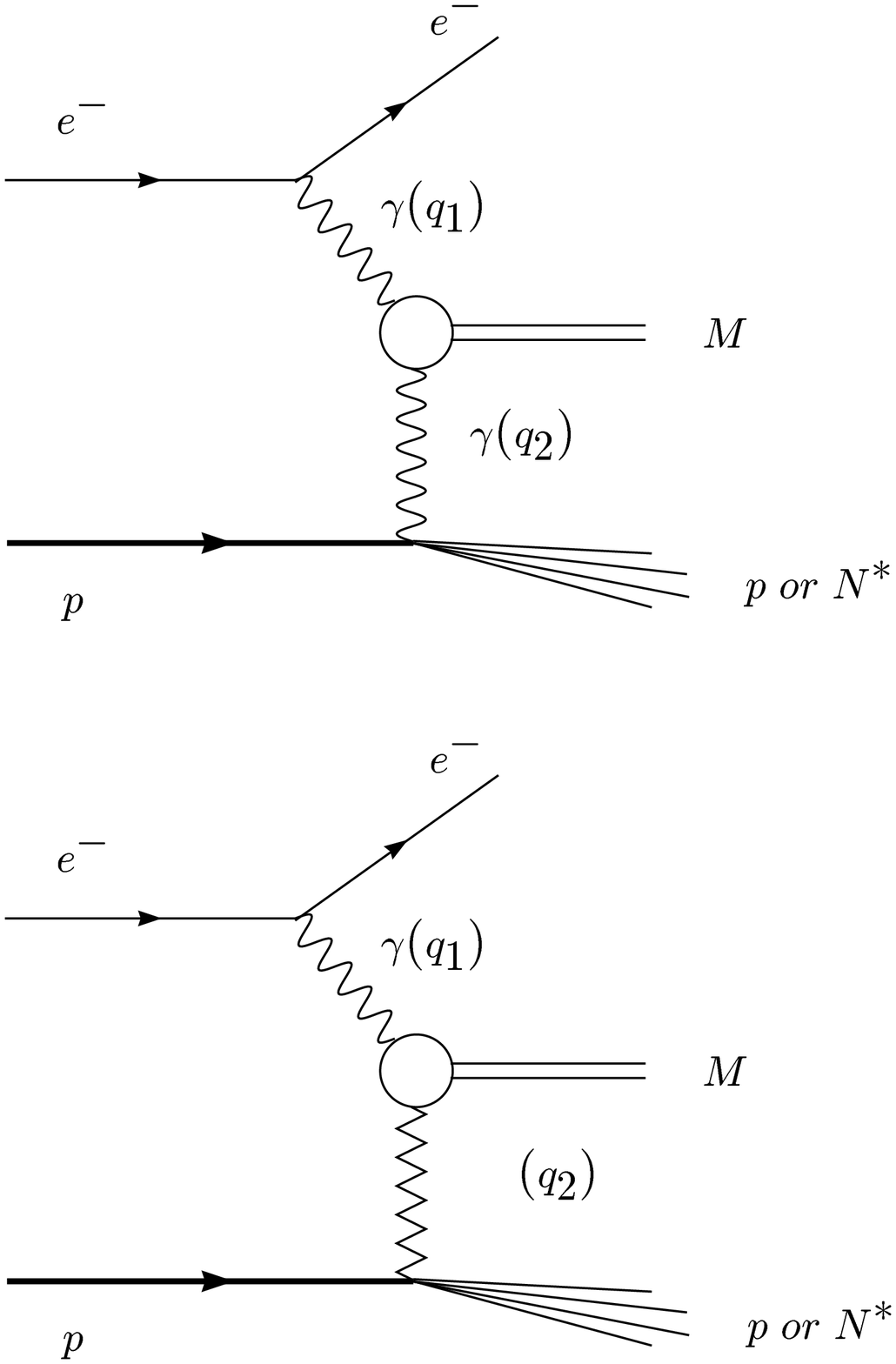,width=0.6\hsize}
  \end{center}
\vspace{-2.85cm}
\hspace{8.1cm}
{\Large $\mathbb{O}$}
\vspace{1.912cm}
\label{fig:pic4}
\end{figure}

\begin{center}
  {\bf Fig. 4}
\end{center}

\end{document}